\documentclass[twocolumn,showpacs,preprintnumbers,amsmath,amssymb,prb]{revtex4}

\usepackage{graphicx}
\usepackage{bm}
\usepackage{subfigure}

\begin{document}
%\tightenlines
%\nonum
\draft

\title{Ratchet potential and rectification effect in Majorana fermion SQUID}

\author{Zhi Wang, Qi-Feng Liang, and Xiao Hu}
\address{International Center for Materials Nanoarchitectonics (WPI-MANA)\\
National Institute for Materials Science, Tsukuba 305-0044, Japan}

\date{\today}

\begin{abstract} Motivated by a recent experimental progress in realizing Majorana fermions (MFs) in a heterostructure of a spin-orbit
coupling nanowire and superconductor (V. Mourik \textit{et al.} Science.1222360), we
investigate a SQUID formed by the novel superconductor-nanowire-superconductor Josephson junction which contains MFs and a conventional superconductor-insulator-superconductor junction. It is shown that the critical current of the SQUID is different for the two current
directions. Since the asymmetric Josepshon current forms a ratchet potential for the dynamics of superconducting phase, a
rectification effect is expected when the SQUID is driven by an ac current. These novel properties are expected to be useful for probing the elusive MFs as well as for their dynamics.
\end{abstract}
\pacs{74.50.+r, 85.25.Dq, 71.10.Pm}

\maketitle

\textit{Introduction.--}
Majorana fermions (MFs) in topological superconductors have drawn much attention recently\cite{kitaev,read,ivanov,fu,fu2,sarma,sau,linder,lutchyn,oreg,lutchyn2,alicea,beenakker,hasan,zhang}, due to their potential application in topological quantum computation\cite{kitaev2,sarmarmp,hasan,jiang,bonderson}.  They have been predicted to
exist in a number of systems, including the spin-triplet $p$-wave superconductor\cite{kitaev,read}, superconductor-topological insulator interface\cite{fu}, and semiconductor-superconductor heterostructure\cite{sau,lutchyn,oreg}.

Among all these candidate systems, the one dimensional  nanowire-superconductor heterostructure is of particular interest\cite{lutchyn,oreg}, due to the well developed nanowire manufacturing technique.
It is composed of a spin-orbit coupling nanowire in proximity to an \textit{s}-wave superconductor under a moderate magnetic field, and MFs are expected at the ends of the wire\cite{kitaev}. Actually, in a recent experiment, a nanowire-superconductor device was fabricated with an InSb nanowire connecting to superconductor NbTiN\cite{kouwenhoven}. A zero-bias peak in the differential conductance has been detected, in agreement with the theoretical prediction of the resonant Andreev effect\cite{lee}, which is  considered as a strong evidence for the existence of MFs.

Here we propose to build a SQUID using the MF nanowire Josephson junction successfully fabricated in this recent experiment \cite{kouwenhoven} and a conventional superconductor-insulator-superconductor (SIS) Josephson junction, as schematically depicted in Fig.~1. We predict a directionally asymmetric critical current in this SQUID due to the unconventional current phase relation (CPR) involving MFs. This direction-dependent critical current is ubiquitous for both adiabatic and fast MF dynamics thus can provide a supportive evidence for the MF existence.
Since this asymmetric critical current corresponds to a ratchet potential, \textit{i.e.} a periodic potential without reflection symmetry, for the "particle" of superconducting phase
\cite{hanggi}, there appears a novel rectification effect in the SQUID, that is, an ac driving current induces a rectified dc voltage. This rectification effect is not only dominated by the dynamics of phase differences in Josephson junctions, but also strongly influenced by the dynamics of the MF state.

\begin{figure}[tb]
\begin{center}
\leavevmode
\includegraphics[clip=true,width=1\columnwidth]{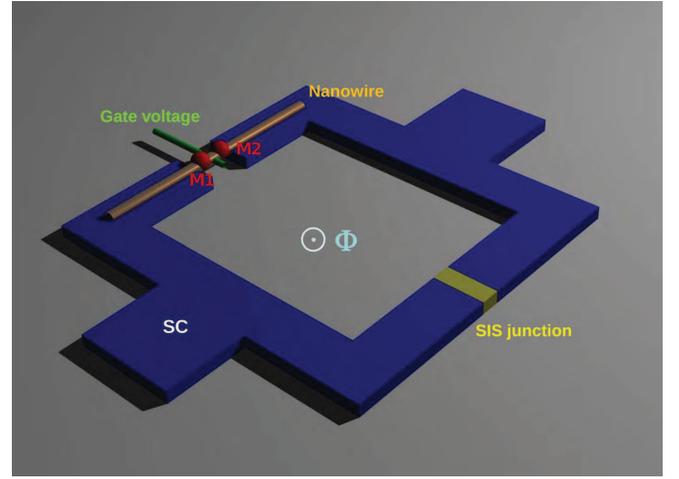}
\caption{Schematic setup of SQUID composed of a novel junction containing two Majorana fermions
and a conventional SIS junction. }
\end{center}
\end{figure}

\vspace{3mm}
\textit{Current-phase relation. --} The MF Josephson junction reported in the recent experiment consists of an InSb nanowire in between two NbTiN superconductors, with a magnetic field along the nanowire; two MFs are expected at the two sides of the tunneling barrier induced by the gate voltage at the middle of the nanowire \cite{kouwenhoven}.
This MF Josephson junction can be described by a tunneling Hamiltonian $\hat H=\hat H_0+\hat H_T$. Here $\hat H_0$ is the Kitaev model for two semi-infinite topological superconducting wires\cite{kitaev},
\begin{eqnarray}
&&\hat H_{0}=  \hat H_L + \hat H_R \\ \nonumber
&& =\sum^{-1}_{j=-\infty} i |\Delta| \gamma_{j, B} \gamma_{j+1,A}
+  \sum^{\infty}_{j=1} i |\Delta| \gamma_{j, B} \gamma_{j+1,A},
\end{eqnarray}
where $|\Delta|$ is the superconducting gap, $\gamma_{j,A}= e^{-i\frac{\varphi_j}{2}} c^{\dagger}_j +e^{i\frac{\varphi_j}{2}} c_j  $,  $\gamma_{j,B}= i(e^{-i\frac{\varphi_j}{2}} c^{\dagger}_j-e^{i\frac{\varphi_j}{2}} c_j  )$ are the Majorana operators, with $c^{\dagger}_j$ the spinless electron creation operator, $\varphi_j=\varphi_L$ for $j\leq 0$, and $\varphi_j=\varphi_R$ for $j > 1$ are the superconducting phases of the two superconductors. The two end MFs $\gamma_{0,B}$ and $\gamma_{1,A}$ do not appear in the above Hamiltonian, which implies a two-fold degenerate ground state for $\hat H_0$. The two superconductors are connected through the electron tunneling,
\begin{eqnarray}
\hat H_T=  T_e c^{\dagger}_L c_R +h.c.,
\end{eqnarray}
where $c_L=c_0$,  $c_R=c_1$, with $T_e$ the electron tunneling matrix.

The Josephson current can be evaluated by the charge change in the left superconductor,
 \begin{eqnarray}
I(t)&&= - \frac{ei}{\hbar}\langle  [\hat H_T,{ \hat N_L}]\rangle =  - \frac{ei}{\hbar} \langle  T_e c^{\dagger}_L c_R -h.c. \rangle \nonumber \\\
&&= \frac{e T i}{\hbar}\langle G| [ (\gamma_{0,B}\gamma_{1,A} -\gamma_{0,A}\gamma_{1,B} ) \sin({\theta}/{2}) \nonumber\\\
&&-  (\gamma_{0,A}\gamma_{1,A}+\gamma_{0,B}\gamma_{1,B} ) \cos ({\theta}/{2})]|G\rangle,
\end{eqnarray}
where $T=T_e/2$, $ \hat N_L$  is the electron number operator of the left superconductor, $\theta=\varphi_R-\varphi_L$ is the phase difference between the right and left superconductors, and $|G\rangle\ $  is the wave function of the ground state.

 For $T<<|\Delta|$, the Josephson current can be calculated by the perturbation method. Since the ground state of the unperturbed Hamiltonian $\hat{H}_0$ is two-fold degenerate, to the lowest-order perturbation, the wave function $|G \rangle \cong  |G_0 \rangle$ is contributed merely from the ground state subspace of $\hat{H}_0$. Therefore, Eq. (3) reads,
 \begin{eqnarray}
I=
 J_M \sin({\theta}/{2})  \langle G_0| i\gamma_{L}\gamma_{R} |G_0\rangle,
\end{eqnarray}
with $\gamma_L =\gamma_{0,B}$, $\gamma_R =\gamma_{1,A}$, and the critical current $J_M = e  {T}/{\hbar}$. The  lowest-order perturbed state $|G_0 \rangle$ evolves with time according to the Schr\"{o}dinger equation,
\begin{eqnarray}
i \hbar \frac {d} {dt} |G _0\rangle = \left[{T} \cos ({\theta}/{2}) (-i\gamma_{L}\gamma_{R})  \right] |G_0 \rangle,
\end{eqnarray}
where only one term in $\hat H_T$ expressed by MF operators has been registered, since other three terms project $|G_0 \rangle$ out of the subspace of degenerate ground states of $\hat H_0$.

In more general cases, the end MFs may slightly deviate from the zero energy due to the correlation to other parts of the superconductors they attached\cite{beenakker,fu2}. This small coupling can be taken into account generally by introducing two terms $\delta_L \gamma_L + \delta_R \gamma_R $ into the Hamiltonian $\hat H_0$,  with $\delta_L$ and $\delta_R$ including all possible interactions to $\gamma_L$ and $\gamma_R$ respectively. As far as MF behaviors are observed, they should be small, and we take
$\delta_L=\delta_R=\delta<<T$ in the present work. Since these two terms account for possible interactions inside each of the two superconductors, they do not change the current expression (4). The dynamic of the end MFs is however enriched, and Eq.~(5) is modified into,
\begin{eqnarray}
i\hbar  \frac {d} {dt} |G_0\rangle=\left[{T} \cos ({\theta}/{2})  (-i\gamma_{L}\gamma_{R})+ \delta(\gamma_L+\gamma_R)\right] |G_0 \rangle.
\end{eqnarray}
The ground state of $\hat H_0$ can be described by the two eigenstates of the MF operators $-i\gamma_{L}\gamma_{R} |\pm \rangle=\pm |\pm \rangle$ as $|G_0 \rangle = P_1 |+\rangle + P_2 |-\rangle$. In this basis, Eq.~(6) takes a 2x2 matrix form,
 \begin{eqnarray}\label{eq1}
  {i \hbar } \frac {d} {dt} \left(\begin{array}{cc}  P_1  \\\  P_2 \end{array} \right)=\left(\begin{array}{cc}  {T}\cos ({\theta}/{2}) &\delta-i\delta \\ \delta+i\delta & -{T}\cos ({\theta}/{2}) \end{array}\right)  \left(\begin{array}{cc}  P_1 \\\  P_2 \end{array} \right),
\end{eqnarray}
and the Josephson current is
 \begin{eqnarray}
I_M(\theta,P_1,P_2)=
 J_M   (|P_2|^2-|P_1|^2)\sin({\theta}/{2}).
\end{eqnarray}

We notice that if the coupling $\delta$ equals zero, Eq. (7) is reduced back to the diagonal one in Eq. (5), and eigenstates $(P_1,P_2)$ are $(1,0)$ or $(0,1)$ corresponding to the two different parities. This leads to a $4\pi$-period CPR known as the fractional Josephson effect\cite{kitaev,lutchyn, alicea, lee2},
 \begin{eqnarray}
 I_M (\theta)=\pm J_{M} \sin({\theta}/{2}),
 \end{eqnarray}
as seen in Fig.~ 2a. With the coupling term $\delta$, the parity is no longer a conserved quantity, and the evolution of $|G_0 \rangle$ becomes a Landau-Zener problem\cite{nori}, which has been studied in many other systems and exhibits rich dynamics. There are two limits in
the dynamics governed by Eq.~(7): For the fast process in which
the phase difference changes quickly, the coupling $\delta$ has no much influence and the system behaves essentially in the same way
as for $\delta=0$. However, for the slow adiabatic process, the instant ground state of the 2x2 matrix is always reached, and the Josephson current is given by diagonalizing Eq.~(7),
 \begin{eqnarray}
I_M (\theta) = J_M  A(\theta) \sin(\theta/2),
\end{eqnarray}
where
\begin{eqnarray}
 A(\theta) =\frac{ 2\cos(\frac{\theta}{2}) \sqrt{\cos^2(\frac{\theta}{2})+\frac{2\delta^2}{T^2}} - 2\cos^2(\frac{\theta}{2}) }{ 2\cos^2(\frac{\theta}{2}) -2\cos(\frac{\theta}{2}) \sqrt{\cos^2(\frac{\theta}{2})+\frac{2\delta^2}{T^2}} +\frac{4\delta^2}{T^2}}.
\end{eqnarray}
As seen in Fig.~2a, the CPR is of $2\pi$ period in this case\cite{fu2,lutchyn,beenakker}. It is clear from Eq.~(8) that the CPR of the MF Josephson junction strongly depends on the dynamics of the end MFs, and generally deviates from the simple sinusoidal function of phase difference realized in conventional Josephson junctions.

\begin{figure}[t]
\begin{center}
\leavevmode
\includegraphics[clip=true,width=1\columnwidth]{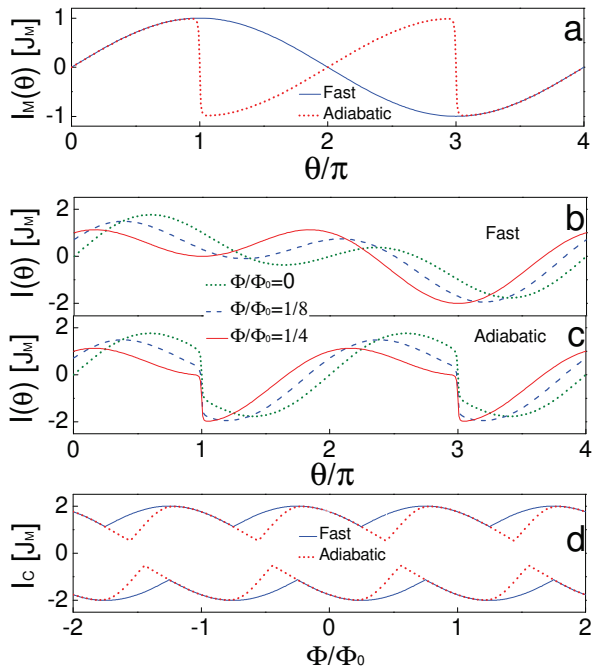}
\caption{(Color online). (a) CPR of MF Josephson junction for the fast process (solid line) and adiabatic process (dotted line). CPR of the SQUID for the fast process (b) and adiabatic process (c) with typical applied magnetic flux $\Phi=0$ (dotted line), $\Phi_0/8$ (dashed line), and $\Phi_0/4$ (solid line).  (d) Interference pattern for the critical current of the SQUID in the fast process (solid line) and adiabatic process (dotted line). Parameters are taken as $\delta=0.01T$ and $J_N=J_M$.}
\end{center}
\end{figure}

\vspace{3mm}
\textit{Asymmetric critical current. --}
 We connect the MF Josephson junction to a SIS Josephson junction, and form a SQUID schematically displayed in Fig. 1. The CPR of the SIS junction is the conventional one $I_N=J_N \sin \theta_N$, where $J_N$ is the critical current and $\theta_N$ is the phase difference.
With a magnetic flux $\Phi$ in the SQUID, the phase differences in the two junctions are related by $\theta_N=\theta+{2 \pi \Phi}/{\Phi_0}$, where $\Phi_0=\hbar/2e$ is the flux quantum. The Josephson current of the SQUID is
 \begin{eqnarray}
I (\theta) = J_N \sin  \left(\theta+\frac{2 \pi \Phi}{\Phi_0}\right)+I_M  ({\theta}).
\end{eqnarray}
 We plug Eqs. (9) and (10) into the above equation, and obtain the CPR of the SQUID for the fast process and adiabatic process respectively.
 The results for three typical $\Phi$ values are shown in Figs.~2b and 2c.

 It is interesting to observe that, for a finite magnetic flux, the currents in the opposite directions, i.e. $I>0$ and $I<0$, are different in both fast and adiabatic processes. As an experimentally observable quantity, we evaluate the magnetic flux dependence of the critical Josephson current. As displayed in Fig.~2d, the critical Josephson currents are different in the two opposite directions for a given magnetic flux, in contrary to the well-known interference pattern in a conventional SQUID.

The asymmetric CPR is caused by the sub-harmonic function $\sin(\theta/2)$ in Eq.~(9) for the fast process, and the non-sine function in Eq.~(10) for the adiabatic process, both originated from the existence of MFs as discussed above. It is due to point out that a SQUID formed by two conventional Josephson junctions cannot realize this asymmetric CPR even if they possess different critical currents. Observed both for fast and adiabatic dynamics, the asymmetry of critical Josephson current with respect to the flowing direction is ubiquitous for MF systems, as compared with the fractional Josephson effect which is expected only for fast dynamics.

The critical current asymmetry in this SQUID structure can serve as an ideal probe for the end MFs. From the experimental perspective, the MF Josephson junction has already been fabricated and its $I-V$ curve measured\cite{kouwenhoven}. Our setup can be built simply by connecting it to an additional SIS junction and forming a SQUID, which is not a difficult task in experiments.
Meanwhile, the measurement of critical current in junction systems is already a routine procedure, leaving no difficulty in experimentally detecting the directional asymmetry.

\vspace{3mm}
\textit{Rectification effect. --}
The asymmetry in critical currents for two flowing directions in the MF SQUID revealed above has
an interesting consequence known as the rectification effect under an ac driving current,
which has been discussed in various contexts so far \cite{hanggi,hanggirmp,wambaugh,sterck}. In the
present MF system, the "particle" of superconducting phase feels a ratchet potential as seen in Fig.~2 due
to the unconventional CPRs in Eqs.~(9) and/or (10), where the translational anti-symmetry
$I(\theta+\pi)=-I(\theta)$ enjoyed by conventional SQUIDs is broken\cite{quintero}.
When the amplitude of the ac driving current
is chosen in between the two different critical currents, pure supercurrent flow is possible only in
one direction, while the current in the opposite direction exceeds the
critical current and contains a normal part. In this case, a rectified voltage is induced across
the junction. As the degree of asymmetry in critical current is different for the adiabatic and
fast processes, this rectification effect is strongly influenced by the MF dynamics.

To be specific, we adopt the resistively and capacitively shunted junction (RCSJ) model, with the phase difference $\theta$ evolves with time according to the dynamic equation,
 \begin{eqnarray}\label{eq2}
 && \frac{ \hbar C}{2e} \frac{d^2\theta}{dt^2} +  \frac{\hbar}{2eR} \frac {d\theta}{dt} - {I(t)} \\\
 && = - J_N  \sin\left(\theta+ \frac{2 \pi \Phi}{\Phi_0}\right)
- J_M (|P_2|^2-|P_1|^2)\sin({\theta}/{2}),  \nonumber
\end{eqnarray}
where $I(t)$ is the external current, and $R$ and C are the resistance and capacitance of the SQUID. The dynamics of the SQUID can be obtained by solving Eqs.~(7) and (13) simultaneously.
The resistance of the SQUID is taken as $R=2 {\rm k}\Omega$, in the same order of that for the MF Josephson junction in
the recent experiment \cite{kouwenhoven}. As a small hysteresis loop in the $I-V$ characteristics was observed experimentally\cite{kouwenhoven}, we focus here on a slightly underdamped case with $Q$-factor $Q\equiv eR\sqrt{2CT}/\hbar=\sqrt{2}$.

\begin{figure}[t]
\begin{center}
\leavevmode
\includegraphics[clip=true,width=0.95\columnwidth]{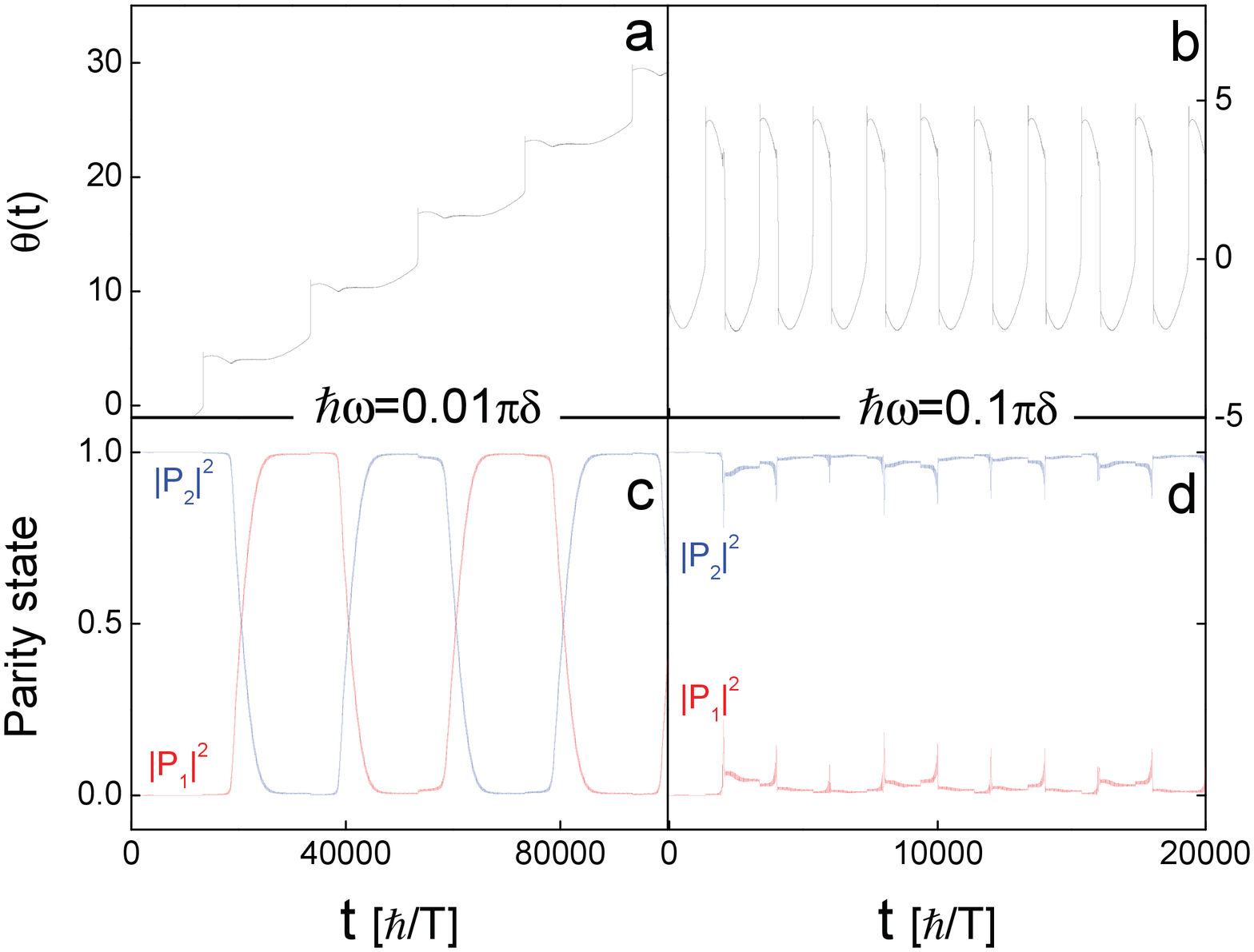}
\caption{(Color online). Dynamics of the phase difference $\theta (t)$ (a,b), and the MF state $|G_0(t)\rangle$ (c,d), under an ac driving current $J_E=J_M$, with driven current frequency $\hbar\omega= 0.01\pi\delta$ (a,c), and $\hbar\omega= 0.1\pi\delta$ (b,d). Other parameters are taken as $J_N=J_M$, $R=2k\Omega$, $Q=\sqrt{2}$ (see text for detailed discussions), and $\delta=0.01T$.}
\end{center}
\end{figure}

\begin{figure}[tb]
\begin{center}
\includegraphics[clip=true,width=1\columnwidth]{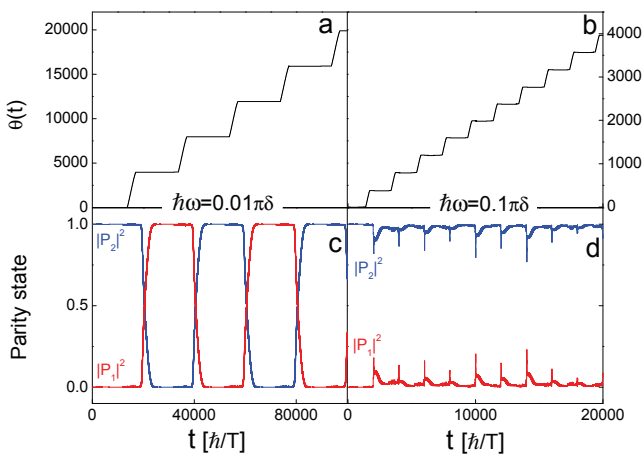}
\caption{(Color online). Same as in Fig. 3 except for $J_E=1.5J_M$. }
\end{center}
\end{figure}

Without loosing generality, let us put $J_{\rm N}=J_{\rm M}$, and take $\Phi=\Phi_0/3$ as an example, where for the adiabatic process the left and right flowing critical currents are $0.866J_M$ and $-1.77J_M$ respectively, while for the fast process, they are $1.37J_M$ and $-1.97J_M$ respectively, as seen in Fig.~2d.

Now we apply an ac driving current $I(t)=J_E \sin (\omega t)$, with $J_E=J_M$.
The time evolutions of the phase difference $\theta(t)$ and the MF state $|G_0(t)\rangle$ are shown in Fig. 3, for two typical frequencies $\hbar\omega  = 0.01\pi\delta$ and $0.1\pi\delta$. It is clear that for $\hbar\omega   =0.01 \pi \delta$, the MF state evolves away from the initial state and takes the instant ground state, a typical adiabatic behavior. Simultaneously, we find that the phase difference increases with time monotonically, which indicates a dc voltage. This is the voltage rectification effect expected from the asymmetric critical current.
For $\hbar\omega = 0.1\pi\delta$, however, the MF state stays at the initial state, a typical fast non-adiabatic process behavior, and the phase difference is oscillating around zero indicating no rectified voltage in the system. The different rectification behaviors for different driving current frequencies come from the value $J_E$, which is in between the two critical currents of different directions for the adiabatic limit, but below the two critical current for the fast process.
For $J_E=1.5 J_M$, which is in between the two critical currents in both limits, the time evolution of the phase difference and the MF state is shown in Fig. 4. We observe the rectification effect for both frequencies.  With $J_M\cong40nA$ suggested from the experiment \cite{kouwenhoven}, the rectified voltage $V=\langle \dot{\theta}\rangle \hbar/2e \cong 0.2 \hbar J_M/2e^2$ is in order of $10 \mu V$.

We numerically simulate various junction parameters, and find this rectification in a large parameter space. The experimental observation of this rectification effect is obviously helpful for identify the existence of MFs. Meanwhile, since it
disappears when the driven frequency approaches $\delta$ under appropriate driving currents, it is possible to draw information on the MF coupling. The rectification effect is scalable, and the rectified voltage can be amplified simply by connecting several SQUIDs in series, making its experimental observation and potential application plausible.

\vspace{3mm}
\textit{Conclusion. --} In summary, we have demonstrated that Majorana fermions produce unconventional current-phase relations in a
Josephon junction including a spin-orbital coupling nanowire, with the detailed forms depending on the dynamics of the Majorana fermions. The SQUID formed by this novel Josephson junction and a conventional one shows ubiquitously direction-dependent critical currents. This ratchet potential established by Majorana fermions for the dynamics of superconducting phase yields a rectified dc voltage across the SQUID when it is driven by an ac current. These novel properties can be explored as useful probes to the elusive Majorana fermions.

This work was supported by WPI Initiative on Materials
Nanoarchitectonics, MEXT, Japan, and partially by Grants-in-Aid for
Scientific Research (No.22540377), JSPS, and CREST, JST.

\end{document}